\documentclass{ws-ijgmmp}
\begin{document}

\markboth{ S.Capozziello, C.A. Mantica, L.G. Molinari}
{Perfect-Fluids in $f(R)$ gravity}

%
\catchline{}{}{}{}{}
%

\title{COSMOLOGICAL PERFECT-FLUIDS  IN $f(R)$  GRAVITY}

\author{SALVATORE CAPOZZIELLO}
\address{
Dipartimento di Fisica ``E. Pancini", Universit\'a di Napoli
	\textquotedblleft{Federico II}\textquotedblright, Napoli, Italy,\\
INFN Sez. di Napoli, Compl. Univ. di Monte S. Angelo, Edificio G, Via Cinthia, I-80126, Napoli, Italy,\\
Tomsk State Pedagogical University, ul. Kievskaya, 60, 634061 Tomsk, Russia.\\
\email{capozziello@na.infn.it}
}

\author{CARLO ALBERTO MANTICA}
\address{I.I.S. Lagrange, Via L. Modignani 65, 
I-20161, Milano, Italy \\
and INFN sez. di Milano,
Via Celoria 16, I-20133 Milano, Italy\\
\email{carlo.mantica@mi.infn.it}
}

\author{LUCA GUIDO MOLINARI}
\address{Dipartimento di Fisica ``A. Pontremoli'',
Universit\`a degli Studi di Milano\\ and INFN sez. di Milano,
Via Celoria 16, I-20133 Milano, Italy\\
\email{luca.molinari@unimi.it}
}

\maketitle

\begin{history}
\received{(Day Month Year)}
\revised{(Day Month Year)}
\end{history}

\begin{abstract}
We show that an $n$-dimensional generalized Robertson-Walker (GRW) space-time with divergence-free conformal 
curvature tensor exhibits a perfect fluid stress-energy tensor for any $f(R)$ gravity model.
Furthermore we prove that a conformally flat GRW space-time is still a perfect fluid in both $f(R)$ and quadratic gravity where other curvature invariants are considered.
\end{abstract}
\keywords{cosmology; extended theories of gravity; perfect fluids; conformal symmetry\\ \\
AMSC: 83C40, 83C05, 83C10.}
\section{Introduction}
Perfect fluids play a crucial role in General Relativity  being the natural sources of Einstein's field equations compatible with  the Bianchi identities.
Thanks to this feature, any source of field equations that can be recast in a perfect fluid form is suitable, in principle, for  solving dynamics having  a well 
posed formulation of the related Cauchy problem 
\cite{brouath, vignolo1,vignolo2}. In cosmology, perfect fluids can represent, at least in a coarse-grained picture, the effective behavior of Hubble flow ranging from inflation to dark energy epochs \cite{Bam12}. 
For these reasons, compatibility of perfect-fluid solutions with modified or extended theories of gravity is a crucial issue to be investigated. 

In this paper, we want to face this problem for the most straightforward generalization of General Relativity which is $f(R)$ gravity, where a generic function of the Ricci scalar is considered in the Hilbert-Einstein action of gravitational field. 

To start these considerations, let us take into account 
a Lorentzian manifold of dimension $n$ whose Ricci tensor has the form 
\begin{align} 
R_{kl} = \frac{R-n\xi}{n-1} u_k u_l + \frac{R-\xi}{n-1} g_{kl} \label{Ricci}
\end{align}
where $g_{kl}$ is the metric, $R=R^k{}_k$ is the curvature scalar, $u_k$ is a given time-like vector field $u_ju^j=-1$, $\xi $ is the eigenvalue $R_{ij}u^j=\xi u_i$. 
The space-time is named {\it perfect fluid space-time}, while in the geometric literature it is known as a quasi-Einstein manifold, with metric of arbitrary signature \cite{ChMa00,[12]}. The reason is that,
by the Einstein's field equations 
\begin{align} 
R_{kl} -\tfrac{1}{2} g_{kl} R = \kappa T_{kl} \label{(1.16)}
\end{align}
the Ricci tensor \eqref{Ricci} implies the stress-energy tensor of a perfect fluid 
 \begin{equation} T_{kl} = (\mu + p) u_ku_l + p g_{kl} \,.\end{equation}
$\kappa$ is Einstein's gravitational coupling.

As said before, a  generalization of Einstein's theory are the so called $f(R)$ theories of gravitation. They were introduced by Buchdahl in 1970 \cite{Buchdahl70} and gained popularity with the works by Starobinsky on cosmic inflation \cite{Starobinsky80}.  More recently, they gained  interest also as a possible mechanism to explain the today observed cosmic acceleration, often dubbed as dark energy \cite{Capoz02}. 
In general, extensions or alternatives to General Relativity are invoked to address the problem of {\it dark side} of the universe (dark energy + dark matter), instead of searching for new material ingredients  (until now not found)  at fundamental level   \cite{CF08,Faraoni,Oikonomou,CDL11}. In such theories, the scalar $R$ in the gravitational 
action is replaced by a smooth function $f(R)$: 
\begin{align*} 
S = \frac{1}{2\kappa} \int d^n x \sqrt{-g} f(R) + S_m
\end{align*}
$g$ is the determinant of the $n$-dimensional metric and 
$S_m$ is the action of matter fields. Variation with respect to $g_{kl}$ gives, modulo surface terms,  the field equations:
\begin{align} 
f'(R) R_{kl} -\tfrac{1}{2} f(R) g_{kl} + [g_{kl} \nabla^2 -\nabla_k\nabla_l] f'(R)  =\kappa T_{kl} \label{fieldeqs}
\end{align}
where a prime denotes derivative with respect
to $R$. It is easy to check that  the property $\nabla_k T^k{}_l=0$ is preserved for any differentiable $f(R)$.

In this paper, we  shall study the following problem related to Eq.\eqref{fieldeqs}: \\
{\it If $R_{kl}$ has the perfect-fluid form \eqref{Ricci}, 
the presence of the terms $\nabla_k\nabla_l R$ 
and $(\nabla_k R)(\nabla_l R)$, prevents $T_{kl}$ to  describe a perfect fluid.}\\ 
We show that for this to happen with any $f(R)$, the space-time has to be a generalized Robertson-Walker (GRW) space-time with harmonic Weyl tensor (that is $\nabla_m C_{jkl}{}^m=0$). 
In $n=4$ they imply that the space-time has the standard RW metric.
For special choices of $R$ and $\xi$, a quasi-Einstein vacuum solution is possible.

Similar conclusions are obtained for  quadratic theories of gravity  where $f(R)$ is replaced by a scalar expression quadratic in the Riemann
tensor and its contractions \cite{DeserTekin03}. In this case,  the space-time must be RW in any space-time dimension.

The paper is organized as follows.
In Section 2 we obtain the conditions for $f(R)$ gravity to admit
a perfect fluid stress-energy tensor with a Ricci tensor of the form \eqref{Ricci}. The  GRW space-times, with null divergence of the Weyl tensor, are discussed in 
Section 3. In Section 4 we give a  lemma on the Hessian (i.e. double covariant 
derivatives) of certain scalar fields in GRW space-times, including the scalar curvature, if
$\nabla_m C_{jkl}{}^m=0$.
In Section 4 we show that a RW space-time gives rise to a perfect
fluid stress-energy tensor in any quadratic gravity theory. Conclusions and outlooks are reported in Section 5.

We adopt the following notations.  For a scalar $S$, we  use  $\dot S = u^m\nabla_m S $ 
(in the frame $u^0=1$, $u^\mu =0$, it is $\dot S = \partial_t S$), $v^2$ for $v^k v_k$
and $\nabla^2$ for $\nabla^k\nabla_k$. The metric tensor has signature ($-,+,\dots,+$).

\section{Conditions for perfect fluids in $f(R)$ gravity}
Let's specify the derivatives $\nabla_k\nabla_l f'(R)$ and $\nabla^2 f'(R)$ in the field equations \eqref{fieldeqs} of $f(R)$ gravity: 
 $\nabla_k\nabla_l f'(R) = f^{'''}(R) (\nabla_k R)(\nabla_l R) + f^{''}(R) \nabla_k\nabla_l R $; transvecting it with $g^{kl}$ gives $\nabla^2 f'(R) = f^{'''}(R) (\nabla_k R)^2 + f^{''}(R) \nabla^2 R$. The field equations become
\begin{align} 
& f'(R)R_{kl}  - [f^{'''}(R) (\nabla_k R)(\nabla_l R) + f^{''}(R) \nabla_k\nabla_l R] \label{fieldeqs2}\\
& + g_{kl} [ f^{'''}(R) (\nabla_k R)^2 + f^{''}(R) \nabla^2 R-\tfrac{1}{2} f(R)]=
\kappa T_{kl} \nonumber
\end{align}
We require: $(\nabla_k R)(\nabla_l R) = ag_{kl} + b u_k u_l$ for some scalar fields $a,b$.
Contraction with $u^l$ gives $\dot R \nabla_k R = (a-b)u_k $ i.e. $\nabla_k R$ is parallel to $u_k$.
Then we must have:
\begin{align}
 \nabla_k R + u_k u^m\nabla_m R =0  \qquad \text{(C1)}
\end{align}
Next, we require: $\nabla_k\nabla_l R = \alpha g_{kl} + \beta u_k u_l$ for some scalar fields 
$\alpha,\beta$. Contraction with $u^l$ gives
$\nabla_k \dot R - (\nabla_k u^l)\nabla_l  R= (\alpha - \beta)u_k$.
The second term is zero by C1: $(\nabla_k u^l)u_l =0$. Then $\nabla_k \dot R = (\alpha - \beta)u_k$.
The derivative of C1 is: 
$\nabla_l \nabla_k R=-\nabla_l (u_k \dot R)= -(\nabla_l u_k)\dot R - (\alpha - \beta)u_k u_l $.
Then, we need the condition
\begin{align}
 \nabla_k u_l = \varphi (g_{kl} +u_k u_l)  \qquad \text{(C2)}
\end{align}
where $\varphi $ is a scalar field, i.e. the time-like unit vector field is ``torse-forming'' \cite{Yano44}. 

Condition C2, together with $u^2 =-1$ and $R_{jk}u^k = \xi u_j$ (implied by \eqref{Ricci}) are the defining properties of a Generalized Robertson-Walker (GRW) space-time. The condition C1 poses
a further constraint on the space-time.

With a perfect fluid Ricci tensor \eqref{Ricci} and conditions C1, C2, we prove that the field equations for $f(R)$ gravity are:
\begin{align} 
 \kappa T_{kl} &= u_ku_l  \left [ \frac{R-n\xi}{n-1} f'(R)  + f^{''}(R) (\varphi \dot R - \ddot R) - f^{'''}(R) (\dot R)^2\right ] \label{fieldeqs3}
 \\
& + g_{kl} \left [ -\tfrac{1}{2} f(R)+\frac{R-\xi}{n-1} f'(R) -[(n-2) \varphi \dot R + \ddot R]f^{''}(R) - f^{'''}(R) (\dot R)^2\right ].
 \nonumber
\end{align}
The stress-energy tensor now describes a perfect fluid with pressure and energy density
\begin{align}
&\kappa p =-\tfrac{1}{2}f(R)+\frac{R-\xi}{n-1}f^\prime(R) -[(n-2)\varphi \dot R + \ddot R]f^{\prime\prime}(R) - (\dot R)^2
f^{\prime\prime\prime}(R) \label{pfR}\\
&\kappa\mu = \tfrac{1}{2}f(R) -\xi f^\prime (R) + (n-1)\varphi \dot R f^{\prime\prime}(R) \label{mufR}
\end{align}
When they both vanish, a vacuum solution results, even in presence of a non-zero Ricci tensor. This means that a cosmological constant term is naturally recovered. This result generalizes the approach often used in cosmology \cite{Capoz02,Lambiase} where a curvature stress-energy tensor is derived to address dark energy and dark matter issues \cite{Cardo}. It is worth noticing that General Relativity is immediately recovered for $f(R)=R$. 
\section{Generalized Robertson-Walker  space-times}
A generalized Robertson-Walker (GRW) space-time  is a 
Lorentzian manifold characterized by the metric \cite{[1]}
\begin{align} 
ds^2 = -dt^2 + q^2(t) g^*_{\mu\nu}(x)dx^\mu dx^\nu \label{GRWmetric}
\end{align}
where $g^*_{\mu\nu}(x)$ is the metric tensor of a Riemannian submanifold $M^*$
of dimension $n-1$ and $q$ is a smooth warping function (or scale factor).
These spaces have been
deeply studied in the past decades by several authors \cite{[8],GutOl09,[25],MaMoDe16,MaMoJMP16,[34],[37],[30],[36]},  see also 
 \cite{GRWSurv17} for a review. 
Recently, B.-Y. Chen gave a covariant characterization in terms of a time-like concircular vector field (\cite{Chen14}, \cite{Chen17} Theorem 4.1). 
An equivalent  one was proven in \cite{GRWSurv17}:\\ 
``A space-time is GRW if and only if there exists 
a time-like unit ($u^2=-1$) and torse-forming (C2) vector field,  
that is also eigenvector of the Ricci tensor''.\\
In the coordinate-frame \eqref{GRWmetric} $u^0=1$, $u^\mu =0$, and: 
\begin{gather}
R_{00} = -(n-1) \frac{\ddot q}{q}, \quad R_{\mu0}=0, \quad R_{\mu\nu}=R^*_{\mu\nu} +g^*_{\mu\nu}[(n-2)\dot q^2 + q\ddot q] \label{riccistar}\\
R = \frac{1}{q^2} R^* + (n-1)\left [2 \frac{\ddot q}{q} + (n-2)\left(\frac{\dot q}{q}\right)^2\right  ]\label{RRstar}
\end{gather}
where $R^*=g^{*\mu\nu}R^*_{\mu\nu}$ is the scalar curvature of $M^*$;
the eigenvalue $\xi $ of the Ricci tensor is the scalar field
\begin{align}
\xi= -R_{ab}u^au^b = (n-1) \frac{\ddot q}{q} \label{xi}
\end{align}
In \cite{MaMoJMP16} the covariant expression of the Ricci tensor in GRW space-times was obtained:
\begin{align}  
R_{kl} = \frac{R-n\xi}{n-1} u_ku_l  + \frac{R-\xi}{n-1} g_{kl} -(n-2) C_{jklm}u^ju^m
\end{align}
where $C_{jklm}$ is the Weyl tensor. In the frame \eqref{GRWmetric} the space components are:
\begin{align}
R_{\mu\nu} = \frac{R-\xi}{n-1}g^*_{\mu\nu} q^2 -(n-2)C_{0\mu\nu 0}\,.
\end{align} 
Using \eqref{riccistar},  it gives: 
\begin{align}
R^*_{\mu\nu} = \frac{R^*}{n-1}g^*_{\mu\nu} - (n-2) C_{0\mu\nu 0} \label{Geba}
\end{align} 
The Ricci tensor has the perfect fluid form \eqref{Ricci} with torse-forming vector field $u_j$ 
(condition C2) if $C_{jklm}u^ju^m =0$. 
We now recall that $u_j$ has the property 
\begin{align}
u_i u^m R_{jklm} + u_j u^m R_{kilm} + u_k u^m R_{ijlm} =0\,,
\end{align} 
 named the Riemann-compatibility \cite{[20]}, 
which implies the Weyl-compatibility \cite{[24]}:
\begin{align}
u_i u^m C_{jklm} + u_j u^m C_{kilm} + u_k u^m C_{ijlm} =0
\end{align}
This shows that $C_{jklm}u^ju^m =0$ if and only if $C_{jklm}u^m=0$. 
The following theorem gives an interesting necessary and sufficient condition:
\begin{theorem}[\cite{MaMoJMP16} Theorem 3.4 and Proposition 3.5] 
On every GRW space-time, with time-like unit torse-forming vector $u_j$, it is 
\begin{align} 
\nabla_m C_{jkl}{}^m =0 \;\Longleftrightarrow \; u_m C_{jkl}{}^m =0\,. 
\end{align}
\end{theorem}
We conclude that condition C2 and the requirement of perfect fluid Ricci tensor are equivalent to the space-time being a 
GRW, with $\nabla_m C_{jkl}{}^m=0$. The next proposition shows that condition C1 is fulfilled:
\begin{proposition}[\cite{MaMoJMP16} Theorem 3.4 and Proposition 3.5] \label{thrm_1.3}
On every GRW space-time, with time-like unit torse-forming vector $u_j$,
if $\nabla_m C_{jkl}{}^m=0$ then:
\begin{align} 
& \nabla_k R + u_k u^m\nabla_m R =0  \label{(1.12)}\\
& [\nabla_i,\nabla_j] R_{kl} = -\frac{\xi}{n-1}[g_{jk}R_{li} - g_{ik}R_{jl} + g_{jl} R_{ik} - g_{il} R_{jk}] 
\end{align}
\end{proposition}
\begin{theorem} 
On an $n$-dimensional GRW space-time with $\nabla_m C_{jkl}{}^m=0$, the stress-energy tensor 
is a perfect fluid in any $f(R)$ theory of gravity.
\begin{proof}
{\rm By hypothesis, the Ricci tensor has the perfect fluid form \eqref{Ricci} and condition C2 holds. 
Condition C1 is also met, by the previous proposition. }
\end{proof}
\end{theorem}

\section{The Hessian and the scalar curvature} 
Let us consider now  the Hessian (second covariant derivatives) of certain scalars
in GRW space-times, and compute the Hessian of the scalar curvature $R$ in the case $\nabla_m C_{jkl}{}^m=0$. 
\begin{lemma}\label{lemmalemma}
If a scalar field $S$ has the property $\nabla_jS+ u_ju^m\nabla_m S=0$, where $u^k$ is
a time-like unit torse-forming vector field, then the Hessian is
\begin{align}
\nabla_j \nabla_k S = Ag_{jk} + B u_j u_k
\end{align}
where $A=-\varphi \dot S$ and $B= -\varphi \dot S + \ddot S$. Then, 
$\nabla^2 S = -(n-1)\varphi \dot S - \ddot S$. 
\begin{proof} \quad $\nabla_j\nabla_k S = -\nabla_j (u_k u^m\nabla_m S)$
\begin{align*}
&= -\varphi h_{jk} u^m\nabla_m S - \varphi u_k h_j{}^m \nabla_m S -u_k u^m\nabla_j\nabla_m S\\
&=-\varphi h_{jk} u^m\nabla_m S  -u_k u^m\nabla_m\nabla_j S\\
&=-\varphi h_{jk} u^m\nabla_m S  +u_k u^m\nabla_m (u_j u^p\nabla_p S)\\
&=-\varphi h_{jk} u^m\nabla_m S  +u_k u_j u^m\nabla_m (u^p\nabla_p S)
\end{align*}
\end{proof}
\end{lemma}
\noindent
The lemma always applies to the scalars $\varphi $ and $\xi $:
\begin{proposition} 
In a GRW space-time, it is
\begin{align}
& \xi = (n-1) (u^m\nabla_m \varphi + \varphi^2)  \label{e1} \\
& \nabla_i \varphi + u_i u^m\nabla_m \varphi =0 \label{e2} \\
& \nabla_i \xi + u_i u^m\nabla_m \xi =0 \label{e3}
\end{align}
\begin{proof}
$R_{jkl}{}^m u_m = [\nabla_j,\nabla_k]u_l = h_{kl} \nabla_j \varphi - h_{jl}\nabla_k \varphi -\varphi^2
(u_j g_{kl}-u_k g_{jl})$. Contraction with $g^{jl}$ gives
$ R_{km} u^m = u_k(u^m\nabla_m \varphi +(n-1)\varphi^2)- (n-2) \nabla_k \varphi $.
If $R_{km}u^m = \xi u_k$, then \eqref{e1} and \eqref{e2} follow.\\
A derivative of \eqref{e1} gives:  $\nabla_i \xi = (n-1) (\varphi h_i{}^m\nabla_m \varphi 
+u^m \nabla_i\nabla_m \varphi+ 2\varphi \nabla_i \varphi)$. The term $h_i{}^m\nabla_m \varphi $
is zero by \eqref{e2}. The next term is: \\
$u^m\nabla_i\nabla_m \varphi = u^m \nabla_m \nabla_i\varphi = - u^m\nabla_m (u_i u^k\nabla_k \varphi) = -u_i u^m\nabla_m (u^k\nabla_k\varphi)$.
Then:  $\nabla_i \xi = - u_i (n-1) (u^m \nabla_m(u^k\nabla_k \varphi ) - 2\varphi u^k\nabla_k\varphi)$
and \eqref{e3} is proven.
\end{proof}
\end{proposition}

\noindent
By \eqref{(1.12)}, the lemma applies to $R$ when $\nabla_m C_{jkl}{}^m=0$, and gives the
expression of the Hessian:
\begin{align} 
\nabla_k \nabla_l R = -\varphi \dot R \, g_{kl} - (\varphi \dot R-\ddot R) u_k u_l\,.
\end{align}
With this expression, the $f(R)$ field equations \eqref{fieldeqs2} take the form \eqref{fieldeqs3} and then the further curvature contribution 
with respect to General Relativity, related to $f(R)$ gravity, can be interpreted as a perfect fluid stress-energy tensor.

\begin{remark}
If $\xi =0$ the covariant divergence of $R_{jk}u^k=0 $ gives $\dot R = -2\varphi R$, and 
\eqref{e1} gives $\dot \varphi +\varphi^2=0$. The Hessian becomes:
$\nabla_k\nabla_l R = 8R\varphi^2 u_ku_l + 2R\varphi^2 g_{kl} $.
Contracting this with $g^{kl}$ gives a Klein-Gordon equation for the scalar curvature,
\begin{align}  
\nabla^2 R = 2 R \varphi ^2 (n-4) \label{(2.5)}\,.
\end{align}
In this sense, the {\it Starobinsky scalaron} is an effective scalar field.\end{remark}

\section{The Friedmann  equations in conformally harmonic GRW space-times}
In view of cosmological applications we note that, in the coordinate
frame \eqref{GRWmetric}, the scalar function $\varphi $ identifies with $\dot q/q $ (see Theorem 2.1 in \cite{[29]})
and the ratio $\dot q/q $ is the Hubble parameter $H$. Then:
\begin{align}
\varphi = \frac{\dot q}{q}=H, \qquad \xi = (n-1)\frac{\ddot q}{q}=(n-1)(\dot H +H^2) \label{Hubble}
\end{align}
Hereafter, we consider GRW space-times with the condition $\nabla_m C_{jkl}{}^m=0$ 
in different theories of gravity.
As a consequence, the Ricci tensor has the form \eqref{Ricci} and $R^*_{\mu\nu}= \frac{1}{n-1} R^* g^*_{\mu\nu}$. 
\subsection{The case of Einstein gravity} The Einstein equations are:
\begin{align}
R_{kl} - \tfrac{1}{2} R g_{kl} = \kappa [ (p+\mu)u_ku_l + p g_{kl}]
\end{align}
The trace and the eigenvalue equation respectively give
$ (n-2)R = -2\kappa p (n-1) + 2\kappa \mu$ and  $2\xi -R= -2\kappa \mu $.
Elimination of  $R$ and $\xi $ by the relations \eqref{RRstar} and \eqref{xi} gives:
\begin{align}
& \kappa \left[ p  +  \mu \frac{n-3}{n-1} \right ]= -(n-2)\frac{\ddot q}{q}  \label{FR1}\\
& \kappa \mu  =  \frac{R^*}{2q^2}  + \frac{1}{2}(n-1)(n-2)\left(\frac{\dot q}{q}\right)^2\,. 
\end{align}
The first is the Raychaudhuri equation for the shear, vorticity, acceleration-free velocity field, i.e. the first Friedmann equation.
The other is the second Friedmann equation. For $n=4$ and defining $q(t)=a(t)$, the scale factor of the universe, the standard cosmological equations are easily recovered. 

\subsection{The case of $f(R)$ gravity} 
By combining Eqs.\eqref{pfR} and \eqref{mufR} for $p$ and $\mu $,  we obtain the analogous of Friedmann Eq. \eqref{FR1}:
\begin{align*}
\kappa\left [p+\frac{n-3}{n-1} \mu \right ] = \frac{Rf'-f}{n-1} -(n-2)\frac{\ddot q}{q} f^\prime 
-\left [\frac{\dot q}{q}\dot R+\ddot R \right ]f^{\prime\prime}- (\dot R)^2 f^{\prime\prime\prime} \nonumber
\end{align*}
It can be written in terms of the Hubble parameter \eqref{Hubble}:
\begin{align}
\kappa\left [p+\frac{n-3}{n-1} \mu \right ] =  \frac{Rf'-f}{n-1} -(n-2)(\dot H +H^2)f^\prime  
- (H\dot R+\ddot R )f^{\prime\prime}- (\dot R)^2 f^{\prime\prime\prime}  \label{RayfR}
\end{align}
Eq. \eqref{mufR} with Hubble's parameter is:
\begin{align}
\kappa \mu & = \tfrac{1}{2}f(R) -(n-1)(\dot H+H^2) f^\prime (R) + (n-1)H \dot R f^{\prime\prime}(R)\nonumber \\
&= \frac{1}{2}(f-Rf^\prime) +\frac{1}{2}\left [\frac{R^*}{q^2} +(n-1)(n-2)H^2\right ] f^\prime+ (n-1)H\, \dot R \, f^{\prime\prime}
\label{faraoni}
\end{align}
where we used $R=2(n-1)\dot H + n(n-1) H^2 + (R^*/q^2)$. 
\begin{remark}
If $R^*=0$, Eq.\eqref{faraoni} coincides with Eq.(75) in  \cite{Faraoni},
Eq. \eqref{pfR} for the pressure is 
\begin{align*}
\kappa p =\tfrac{1}{2}(Rf^\prime -f)-\tfrac{1}{2} (n-2)[2\dot H +(n-1)H^2] f^\prime -[(n-2)H \dot R + \ddot R]f^{\prime\prime} - (\dot R)^2
f^{\prime\prime\prime}
\end{align*}
It coincides with Eq.(76) in \cite{Faraoni} despite of the greater generality of the metric $g^*$.
\end{remark}

\begin{example}
Starobinsky considered the case $f(R)= R+\alpha R^2$ where $\alpha $ is a parameter. The pressure $p$ and the energy density $\mu $
are, in $n=4$ and with $R^*=0$:
\begin{align*}
& \kappa\left (p+\tfrac{1}{3} \mu \right ) = -2(\dot H +H^2)+\alpha[ \tfrac{1}{3} R^2 -4(\dot H +H^2)R - 2(H\dot R+\ddot R )]\,, \\
&\kappa \mu  = 3H^2 +\alpha [-\tfrac{1}{2}R^2 +6H^2 R + 6H \dot R ]\,,\\
& R=6\dot H + 12 H^2\,.
\end{align*} 
This case has been widely studied in literature (see \cite{OrlyFelix} and references therein). 
\end{example}

\section{Conformal tranformations on GRW spacetimes}
The above considerations can be extended by taking into account conformal transformations.
In a conformal transformation \cite{Wald} the metric tensor $g_{ij}$ is replaced by a locally rescaled one, $\bar g_{kl}(x) = e^{2\sigma (x)}g_{kl}(x)$. 
The Weyl $(1,3)$ tensor (also called the {\it conformal tensor})  is invariant, $\bar C_{jkl}{}^m = C_{jkl}{}^m $, while $\bar C_{jklm} = e^{2\sigma} C_{jklm} $.
With conformal rescaling, the Christoffel symbols, the Ricci tensor and the divergence of the Weyl tensor transform as:
\begin{align}
& \bar\Gamma^m_{ij}  = \Gamma^m_{ij} +\delta^m_j \nabla_i\sigma +\delta^m_i \nabla_j\sigma -g_{ij} \nabla^m\sigma \\
& \bar R_{jk} = R_{jk} - (n-2) [\nabla_j\nabla_k\sigma-(\nabla_j\sigma)(\nabla_k\sigma) +g_{jk}(\nabla^p\sigma)(\nabla_p\sigma)] -g_{jk} \nabla^2\sigma
 \label{confRicci}\\
& \bar\nabla_m \bar C_{jkl}{}^m   = \nabla_m C_{jkl}{}^m  + (n-3)  C_{jkl}{}^m\nabla_m \sigma  \label{confnablaC}
\end{align}
%
With these preliminaries, we can enunciate the following
\begin{theorem}
A conformal transformation $\bar g_{ij}=e^{2\sigma}g_{ij}$ with $\nabla_k\sigma = -u_k \dot \sigma $ maps a GRW space-time ($M,g$) to
a GRW space-time ($M,\bar g$).
\begin{proof}
The torse-forming time-like unit vector field $u_k$ of ($M,g$) is rescaled to $\bar u^k = e^{-\sigma} u^k$, so that $\bar g_{ij}\bar u^i\bar u^j=-1$.  It is
$\bar u_k =e^\sigma u_k$.
 With $\nabla_k\sigma = - u_k \dot \sigma $ let's evaluate:
\begin{align*}
\bar\nabla_i \bar u_j =& e^\sigma ( u_j \nabla_i\sigma + \varphi g_{ij} + \varphi u_iu_j - (\nabla_i\sigma) u_j - (\nabla_j\sigma) u_i + g_{ij}u_m\nabla^m\sigma)\\
=& e^\sigma ( \varphi e^{-2\sigma} \bar g_{ij} +\varphi u_iu_j + u_i u_j \dot\sigma + g_{ij} \dot\sigma)\\
=& e^{-\sigma} (\varphi +\dot\sigma ) (\bar g_{ij} + \bar u_i \bar u_j )
\end{align*}
Therefore $\bar u_k$ is torse-forming in ($M,\bar g$), with $\bar\varphi = e^{-\sigma } (\varphi + \dot\sigma )$. 
To check that it is an eigenvector of $\bar R_{jk}$, let us note that by Lemma \ref{lemmalemma}: 
$\nabla_j\nabla_k\sigma = -\varphi (u_ju_k+g_{jk}) \dot\sigma + u_ju_k\ddot\sigma $. Then $\nabla^2\sigma = -(n-1)\varphi\dot\sigma -\ddot\sigma $,
and we are ready to evaluate 
\begin{align}
\bar R_{ij} =  R_{ij} +(n-2) [ \varphi\dot\sigma +\dot \sigma^2 -\ddot\sigma] u_i u_j +[(2n-3)\varphi \dot\sigma + (n-2) \dot\sigma^2 +\ddot\sigma]  g_{ij}  \label{confRicciperf}
\end{align}
then $\bar u^j =e^{-\sigma} u^j$ is eigenvector of $\bar R_{ij}$, and ($M,\bar g$) is a GRW space-time.
\end{proof}
\end{theorem}
From \eqref{confRicciperf} we obtain the curvature scalar, the eigenvalue, and the Einstein tensor:
\begin{align}
& \bar R=e^{-2\sigma}[ R+2(n-1)^2\varphi\dot\sigma + (n-1)(n-2) \dot\sigma^2+2(n-1) \ddot \sigma ]\\
& \bar\xi =e^{-2\sigma} [\xi +(n-1)(\varphi \dot\sigma + \ddot\sigma )]\\
& \bar R_{ij}-\tfrac{1}{2}\bar R \bar g_{ij} = R_{ij}-\tfrac{1}{2} R g_{ij} +(n-2)(\varphi \dot\sigma +\dot\sigma^2-\ddot\sigma) u_iu_j  \label{Einst}\\
& \qquad\qquad\qquad  - (n-2) [ (n-2)\varphi\dot\sigma +\tfrac{1}{2}(n-3)\dot\sigma^2 +\ddot\sigma ] g_{ij}\nonumber
\end{align}
\begin{proposition}\label{prop_6_2}
Consider a GRW space-time with 
$\nabla_m C_{jkl}{}^m =0$. A conformal transformation $\bar g=e^{2\sigma}g$ with 
$\nabla_k\sigma = -u_k \dot \sigma $ maps it to a GRW space-time with  
$\bar\nabla_m \bar C_{jkl}{}^m =0$.
\begin{proof}
For a GRW space-time the conditions $\nabla_m C_{jkl}{}^m = 0$ and $u_m C_{jkl}{}^m =0$ are equivalent. Therefore, by Eq.\eqref{confnablaC},  if
 $\nabla_m C_{jkl}{}^m = 0$ then also $\bar \nabla_m \bar C_{jkl}{}^m = 0$.
\end{proof}
\end{proposition}
\begin{remark}
If $(M,g)$ is a GRW and $\nabla^m C_{jklm}=0$, then the Ricci tensor is a perfect fluid. A conformal map with $\nabla_k\sigma = -u_k\dot\sigma $
gives a GRW space-time $(M,\bar g)$ with perfect fluid Ricci tensor
$$ \bar R_{ij} = \frac{\bar R-n\bar\xi}{n-1}\bar u_i\bar u_j + \frac{\bar R- \bar\xi}{n-1}\bar g_{ij} $$
\end{remark}
\noindent
Considering $f(R)$ gravity,  one exploits the conformal map $\bar g_{kl} =e^{2\sigma} g_{kl}$ \cite{Barrow88}, with
\begin{align}
\sigma = \frac{1}{n-2}\log [f'(R)]  \label{conf_fprime}
\end{align}
and $f'(R)>0$, to map $f(R)$  gravity to Einstein gravity minimally coupled 
to an extra scalar field. For such transformation, the space-times
$(M,g)$ and $(M,\bar g)$ are named {\it Jordan} and {\it Einstein frame} respectively.

If $(M,g)$ is a GRW space-time with $\nabla^m C_{jklm}=0$, the transformation \eqref{conf_fprime} 
satisfies the hypothesis of Prop.\ref{prop_6_2}: it
is ${\displaystyle \nabla_k\sigma = \frac{1}{n-2}(f''/f')\nabla_k R}$. The condition $\nabla_m C_{jkl}{}^m=0$
implies $\nabla_k R = -u_k \dot R$ and therefore $\nabla_k\sigma = -u_k\dot\sigma $, with ${\displaystyle \dot\sigma = \frac{1}{n-2}(f''/f')\dot R }$.
It is also ${\displaystyle \ddot\sigma = \frac{1}{n-2}[(f''/f')\ddot R +(f''/f')' \dot R^2]}$. 
The Einstein tensor in the Einstein frame is obtained from \eqref{Einst}:
\begin{align*}
&\bar R_{ij} - \tfrac{1}{2}\bar R \bar g_{ij} =\frac{1}{f'} \left [ f' \frac{R-n\xi }{n-1} + f''(\varphi \dot R -\ddot R) -f''' \dot R^2
+\frac{n-1}{n-2}\frac{(f''\dot R)^2}{f'}  \right  ] u_iu_j  \\
&  + \frac{1}{f'} \left [f'\frac{R-\xi}{n-1}-\frac{1}{2}f' R-f''[(n-2)\varphi \dot R +\ddot R] -f'''\dot R^2 +\frac{n-1}{2(n-2)}\frac{(f''\dot R)^2}{f'} \right ] g_{ij}  \nonumber
\end{align*}
Comparison with the equation of motion in the Jordan frame \eqref{fieldeqs3}, gives:
\begin{align}
\bar R_{ij} - \tfrac{1}{2}\bar R \bar g_{ij} =& \frac{\kappa}{f'} T_{ij}+ \frac{n-1}{n-2}\frac{(f''\dot R)^2}{f'^2}  (u_iu_j  +\tfrac{1}{2}g_{ij})
-\tfrac{1}{2}(R-\frac{f}{f'})  g_{ij} \,. 
\end{align}
In this sense, results for perfect fluids in the Jordan frame can be transformed to the Einstein frame and back.

%
%
\section{Robertson-Walker in quadratic gravity}
A special consideration deserves quadratic gravity, that is Hilbert-Einstein action corrected with quadratic combinations of curvature invariants \cite{DeserTekin03}. It is based on the following integral action
\begin{align}  
S=\int d^n x \sqrt{-g} \Big[ \frac{R-2\Lambda_0}{\kappa} + \alpha R^2 + \beta R_{ij} R^{ij}
 + \gamma (R_{jklm} R^{jklm} - 4R_{jk}R^{jk} + R^2)\Big] + S_m  \nonumber
\end{align}
As first remark, we  note that the term 
$ {\cal G}=R_{jklm} R^{jklm} - 4R_{jk}R^{jk} + R^2 $
is the Gauss-Bonnet topological invariant, whose integral is equal to zero for $n=4$. It can contribute to the cosmological dynamics if functions of it are considered in the so called $f(R,{\cal G})$ gravity (see e.g. \cite{Felix,odintsov}).

Variation of the action with respect to the metric gives the stress-energy tensor:
\begin{align*} 
& T_{kl} = \frac{1}{\kappa} (R_{kl}- \tfrac{1}{2} R g_{kl} +\Lambda_0 g_{kl} ) +
2 \alpha R (R_{kl} -\tfrac{1}{4} R g_{kl}) +
(2\alpha + \beta) (g_{kl} \nabla^2 - \nabla_k \nabla_l)R \nonumber \\
&+ 2\gamma [R R_{kl}-2R_{akbl} R^{ab} + R_{kcde} R_l{}^{cde} -2R_{ka}R_l{}^a -
\tfrac{1}{4} g_{kl} ( R_{jklm}R^{jklm} - 4 R_{kl}R^{kl} + R^2)]\\
& +\beta \nabla^2 (R_{kl}-\tfrac{1}{2} R g_{kl} ) + 2\beta (R_{akbl}-\tfrac{1}{4} g_{kl} R_{ab} ) R^{ab}. \nonumber
\end{align*}
Despite the complicated expression, we are able to prove the following
\begin{theorem} 
On an $n$-dimensional Robertson-Walker space-time the stress energy tensor is a perfect fluid in any quadratic theory of gravity.
\begin{proof}
A RW space-time may be characterised as a GRW space-time with zero Weyl tensor.
Then, the Ricci tensor has the perfect fluid form \eqref{Ricci}, and the Riemann tensor has the
expression (\cite{MaMoJMP16}, Proposition 3.5):
\begin{align}
R_{jklm} =& \frac{1}{(n-1)(n-2)}\Big[ (2\xi -R) (g_{kl}g_{jm} - g_{km} g_{jl}) \\
& + (n\xi-R) ( g_{jm} u_ku_l -
g_{km} u_j u_l+ g_{kl} u_j u_m - g_{jl} u_k u_m) \Big ] \nonumber
\end{align}
Consider the expression of the divergence of the Weyl tensor
\begin{align*}  
\nabla_m C_{jkl}{}^m = \frac{n-3}{n-2} \left [ \nabla_k R_{jl} - \nabla_j R_{kl} + \frac{\nabla_j R g_{kl} - \nabla_k R g_{jl}}{2(n-1)} \right ]
\end{align*}
With $C_{jkl}{}^m=0$, a further derivative gives:
\begin{align*}  
0 =  \nabla^2
R_{kl} -\frac{\nabla^2 Rg_{kl}}{2(n-1)}- [\nabla_j,\nabla_k]R^j{}_l  - \frac{n-2}{2(n-1)} \nabla_k\nabla_l R
\end{align*}
$\nabla_k \nabla_l R$ has the perfect fluid form and, by Theorem \ref{thrm_1.3}, also $[\nabla_j,\nabla_k]R_l{}^j$ has the perfect fluid form. 
It follows that 
$\nabla^2 R_{kl}$ has the perfect fluid form.\\
 $R_{ka}R_l{}^a$ has the perfect fluid form. The expression of the Riemann tensor given above implies 
 that also $R_{akbl}R^{ab}$, $R_{kcde} R_l{}^{cde}$ have a perfect fluid form.
It follows that the stress-energy tensor has the same form.
\end{proof}
\end{theorem}
In other words, we can say that quadratic gravity contributions can be always recast into dynamics as a perfect fluid stress-energy tensor.

\section{Conclusions and outlooks}
Extensions and modifications of General Relativity have a prominent role in addressing the problems of dark energy and dark matter (the so called {\it dark side}). In this perspective,  the shortcoming 
of Einstein's theory to fit astrophysical and cosmological structure at  infrared scale, without huge amounts of exotic fluids, would be in some sense {\it solved} by requesting more degrees of freedom (more ``geometry") to describe the gravitational interaction. Besides, the approach could solve the apparent lack of new ``material" ingredients that, until now, have not been found by fundamental physics experiments. In any case, the General Relativity paradigm is extremely efficient in describing cosmology, then the issue is to model any further contributions under the standard of perfect fluids that act as effective sources in the cosmological equations.\\
With this perspective in mind, it is possible to extend Einstein's theory by considering $f(R)$ gravity and enquiring whether the further degrees of freedom in the gravitational action can be modeled as perfect fluids sourcing the field equations. 

In this paper, we rigorously addressed this question by  demonstrating that any $f(R)$ model can be recast as a cosmological fluid in generalized RW space-times.  This result has been used several times in cosmology \cite{Cardo} but never, to our knowledge, rigorously demonstrated. 

As a general remark, we expect that the results in this paper can be generalized to other geometric corrections to the gravitational action, that can contribute as a perfect fluid in the field equations and be tested by some cosmographic analysis \cite{OrlandoDunsby}. In a forthcoming paper, the present approach will be generalized to other extended gravity theories.
 
\section*{Acknowledgments}
S. C.   acknowledges the support of  INFN ({\it iniziative specifiche} TEONGRAV and QGSKY).
This paper is based upon work from COST action CA15117 (CANTATA), supported by COST (European 
Cooperation in Science and Technology). 


%
\vfill
\end{document}